# Optical Signature of Bias-Induced Carrier Injection in a $WSe_2/MoS_2$ Heterostructure Diode

Yasuyuki Hirata,[a] Hibiki Kinpara,[a] Shinya Ohno,[b] and Yoshihiro Miyauchi[a, †]

[a] *Department of Applied Physics, National Defense Academy of Japan, 1-10-20 Hashirimizu, Yokosuka, Kanagawa 239–8686, Japan*

[b] *Department of Physics, Faculty of Engineering, Yokohama National University, 79-5 Tokiwadai, Hodogaya-ku, Yokohama, 240–8501, Japan*

[†] *Corresponding author: miyauchi@nda.ac.jp*

## Abstract

We investigate the bias-dependent photoluminescence (PL) response of a $WSe_2/MoS_2$ heterostructure diode. Under forward bias, trion-related emission in $WSe_2$ is markedly enhanced, whereas reverse bias produces minimal spectral changes. This asymmetric response cannot be explained by electric-field effects alone and instead reflects carrier injection at the functional interface. These results demonstrate that PL provides a direct optical probe of carrier injection in an operating van der Waals heterostructure. More broadly, this approach establishes spatially resolved PL spectroscopy as a versatile tool for pinpointing the interfaces governing device operation in van der Waals heterostructures.

**Keywords** *$WSe_2/MoS_2$ heterostructure, trion, carrier injection, photoluminescence, van der Waals heterojunction diode*

Two-dimensional (2D) transition-metal dichalcogenides (TMDs) have attracted considerable attention because of their strong Coulomb interactions, layer-dependent band structures, and robust excitonic responses even at room temperature.[1-3] In particular, a monolayer TMD exhibits pronounced neutral exciton and trion features in its photoluminescence (PL), making it a sensitive platform for probing carrier density and charge dynamics.[4-6]

van der Waals heterostructures assembled from different TMD monolayers provide additional opportunities for band engineering at atomically sharp interfaces. Such heterostructures frequently exhibit type-II band alignment, which promotes interfacial charge transfer and spatial separation of electrons and holes.[7–9] Among them, $WSe_2/MoS_2$ heterostructures are known to form type-II band alignment, in which the conduction-band minimum lies in $MoS_2$ and the valence-band maximum lies in $WSe_2$.[7] This band alignment has therefore been widely investigated by optical spectroscopy and transport measurements.[9–11]

Bias-dependent optical responses have also been reported in TMD-based structures, where an applied voltage can modify PL through electric-field effects, charge transfer, or carrier accumulation.[12-14] However, in operating heterostructure devices, it remains difficult to distinguish whether the observed PL modulation arises from electric-field effects or from carrier injection across the heterointerface. Distinguishing between these is essential for linking device operation with optical observables.

Spatially resolved PL spectroscopy has been widely employed as a non-contact technique for investigating local optical properties of two-dimensional semiconductors, including exciton distributions, defects, and carrier dynamics.[15–18] Despite these advances, spatially resolved PL has primarily been utilized for material characterization, and its capability for identifying the functional heterointerface governing device operation under electrical bias remains largely unexplored. Furthermore, in devices containing multiple heterointerfaces and competing current paths, it remains unclear whether spatially resolved PL can selectively identify the junction governing the observed diode operation.

In this work, we investigate a $WSe_2/MoS_2$ heterostructure diode using bias-dependent PL spectroscopy. We show that forward bias markedly enhances a trion-related emission in $WSe_2$, whereas reverse bias produces only minor spectral changes. This asymmetric response indicates that the PL modulation is governed predominantly by carrier injection rather than by electric-field effects alone. These findings demonstrate that bias-dependent PL spectroscopy provides a direct optical probe of carrier injection in an operating van der Waals heterostructure.

The $WSe_2/MoS_2$ heterostructure device was fabricated by stacking an exfoliated $WSe_2$ flake containing both monolayer (ML) and few-layer (FL) regions onto an exfoliated ML-$MoS_2$ flake on pre-patterned comb-shaped gold electrodes, as shown in Fig. 1(a). This geometry forms both ML-$WSe_2$/ML-$MoS_2$ and FL-$WSe_2$/ML-$MoS_2$ vertical heterointerfaces within the same device, together with a lateral FL-$WSe_2$/ML-$WSe_2$ junction, as illustrated in Fig. 1(b).

The band alignment of the ML-$WSe_2$/ML-$MoS_2$ heterostructure under forward bias is schematically depicted in Fig. 1(c). The system exhibits type-II band alignment, in which the conduction-band minimum of ML-$MoS_2$ lies below that of ML-$WSe_2$.[7] At equilibrium, this band offset facilitates electron transfer from ML-$WSe_2$ into ML-$MoS_2$. At the heterointerface, electrons can be exchanged between the layers owing to the type-II band alignment, as commonly observed in TMD heterostructures.[9,10] Under forward bias, however, electrons are injected from ML-$MoS_2$ into ML-$WSe_2$, as illustrated in Fig. 1(c).

To isolate the current flow through the heterostructure, laser ablation was employed to remove excess material and interrupt unintended conduction paths, as illustrated in Fig. 1(a), Fig. 1(b), and Fig. S1(b). Because the laser-affected region extend beyond the visibly damaged area owing to the finite laser spot size, the lateral $WSe_2$ transport path is expected to be substantially suppressed while preserving the ML-$WSe_2$/ML-$MoS_2$ heterointerface. The evolution of the current–voltage (*I–V*) characteristics during this process is depicted in Fig. S2. Prior to modification, the *I–V* response exhibits nearly linear behavior, whereas after removing unwanted conduction paths, a clear rectifying characteristic emerges (Fig. 1(d)), indicating diode-like transport across the ML-$WSe_2$/ ML-$MoS_2$ heterointerface. These results demonstrate that suppressing competing current paths is essential for revealing the intrinsic transport behavior of the ML-$WSe_2$/ML-$MoS_2$ heterointerface.

To confirm the crystallographic orientation and the formation of the ML-$WSe_2$/ML-$MoS_2$ heterostructure, incident polarization-resolved second-harmonic generation (SHG) measurements were performed, as shown in Fig. 2. The SHG intensity exhibits a characteristic angular dependence consistent with the symmetry of monolayer TMDs. From the phase difference of the polarization-resolved SHG patterns, the relative crystal orientation between ML-$WSe_2$ and ML-$MoS_2$ was determined to be 56.5°, indicating a well-defined stacked interface. Similar approaches have been widely used to determine crystal orientation in TMD systems.[19,20]

Having established the stacking configuration from the SHG measurements, we next investigate the carrier dynamics at the heterointerface by bias-dependent PL spectroscopy. These PL measurements were performed on the as-fabricated device, prior to the laser-ablation process used to isolate the current path. Micro-PL measurements were performed using a 405 nm excitation laser with a power of 8 mW, focused onto the sample through a 100× objective lens (NA = 0.95), producing a spot diameter of approximately 2 μm. The PL emission was collected and analyzed using an Ocean Insight SR-series fiber-optic spectrometer. Figure 3 presents the PL spectra measured on the unmodified device under various bias conditions. In the unbiased spectrum shown in Fig. 3(a), the emission is dominated by the neutral exciton ($X^0$), with a lower-energy shoulder attributed to the negatively charged trion ($X^-$). The high-energy peak at ~1.65 eV is assigned to the neutral exciton ($X^0$), consistent with previous room-temperature PL studies of ML-$WSe_2$.[21] The low-energy shoulder at ~1.60 eV is attributed to the negatively charged exciton ($X^-$), consistent with previous room-temperature photoluminescence studies of monolayer $WSe_2$ reporting neutral-exciton and negatively charged trion emissions near 1.64 and 1.59 eV, respectively.[22] This assignment is further supported by electron-doping experiments demonstrating a pronounced redshift of the trion emission with increasing electron density.[23]

Under forward bias, the trion emission is markedly enhanced. Multiple-Lorentzian spectral decomposition reveals that this enhancement is accompanied by a decrease in the $X^0$ contribution and a corresponding increase in the $X^-$ component. By contrast, reverse bias produces only minor spectral variations. This asymmetric response indicates preferential electron injection into the $WSe_2$ layer under forward bias.

Although a momentum-indirect Γ–K interlayer exciton has been reported near 1.6 eV in $MoS_2/WSe_2$ heterostructures,[24] interlayer excitons generally exhibit much smaller oscillator strengths than intralayer excitons because of the spatial separation between electrons and holes.[25] Together with the observed reduction in the $X^0$ contribution, the pronounced enhancement near 1.60 eV is therefore more consistently attributed to bias-induced trion formation than to interlayer-exciton emission.

The enhancement of the $X^-$ emission reflects an increase in the electron population in ML-$WSe_2$, consistent with carrier transfer across the heterointerface. As illustrated by the proposed band alignment in Fig. 1(c), electrons are injected from ML-$MoS_2$ into ML-$WSe_2$ under forward bias, leading to electron accumulation within the ML-$WSe_2$ layer. The proposed band alignment is based on the experimentally determined type-II band alignment of ML-$WSe_2$/ML-$MoS_2$ heterostructures,[7] which is further supported by previous studies on $WSe_2/MoS_2$ heterojunction devices.[26] Such a band alignment promotes interlayer charge transfer, as commonly observed in TMD heterostructures.[9,10] Negatively charged trions are formed by the binding of neutral excitons with excess electrons, and their emission intensity is highly sensitive to the local electron density.[23] Figure S4 schematically illustrates this trion formation process. Under forward bias, the injected electrons promote the conversion of optically generated neutral excitons ($X^0$) into negatively charged trions ($X^-$), consistent with the observed increase in the $X^-$ emission and the corresponding decrease in the $X^0$ contribution.

These results indicate that PL provides a sensitive probe of local carrier injection and that trion emission serves as an optical signature of electron accumulation at the ML-$WSe_2$/ML-$MoS_2$ heterointerface.

To clarify whether the bias-dependent PL modulation originates from the heterointerface or the constituent monolayers, bias-dependent PL measurements were performed on isolated ML-$MoS_2$ and ML-$WSe_2$ regions (Figs. S5 and S6). Both monolayers exhibit negligible spectral changes under forward and reverse bias, indicating that the pronounced trion enhancement observed in Fig. 3 is an intrinsic property of the ML-$WSe_2$/ML-$MoS_2$ heterointerface.

Practical heterostructure devices often contain multiple interfaces and competing current paths. The PL measurement position shown in Fig. S7 contains the ML-$WSe_2$/ML-$MoS_2$ heterointerface together with the FL-$WSe_2$/ML-$MoS_2$ heterointerface and the lateral FL-$WSe_2$/ML-$WSe_2$ junction, making it difficult to identify the functional junction from electrical measurements alone. When PL spectra are measured from this region, the $X^-$ intensity increases by only approximately 1.8 times under forward bias (Fig. S8). In contrast, restricting the measurement to the ML-$WSe_2$/ML-$MoS_2$ heterointerface yields an approximately 18-fold enhancement (Fig. 3), demonstrating that the dominant bias-induced carrier accumulation occurs specifically at the ML-$WSe_2$/ML-$MoS_2$ heterointerface.

To distinguish whether the rectifying behavior originates from the ML-$WSe_2$/ML-$MoS_2$ heterointerface or lateral transport through the $WSe_2$ flake, the lateral $WSe_2$ conduction path was removed by laser ablation while preserving the ML-$WSe_2$/ML-$MoS_2$ heterointerface. Before laser ablation, the current predominantly flowed through the lateral $WSe_2$ region, where the layer-thickness asymmetry was already present but no pronounced rectifying behavior was observed (Fig. S2(b)). After suppressing this competing transport path, clear rectifying *I*–*V* characteristics emerged (Fig. 1(d)), indicating that the layer-thickness asymmetry alone cannot account for the diode operation.

Taken together, the spatially resolved PL measurements identify the ML-$WSe_2$/ML-$MoS_2$ heterointerface as the principal location of bias-induced carrier accumulation. The subsequent laser-ablation experiment verifies that this interface is the functional heterointerface responsible for the rectifying diode operation. More broadly, these results establish spatially resolved PL spectroscopy as a non-destructive approach for identifying functional heterointerfaces in operating van der Waals heterostructure devices.

## SUPPLEMENTARY MATERIAL

See the supplementary material for additional device schematics, I–V characteristics, spectral analyses, control PL measurements, spatially resolved PL measurements, and a schematic illustration of the trion formation mechanism.

## Acknowledgments

This work was supported by JSPS KAKENHI Grant Number JP23K26540.

## Data Availability Statement

The data that support the findings of this study are available from the corresponding author upon reasonable request.

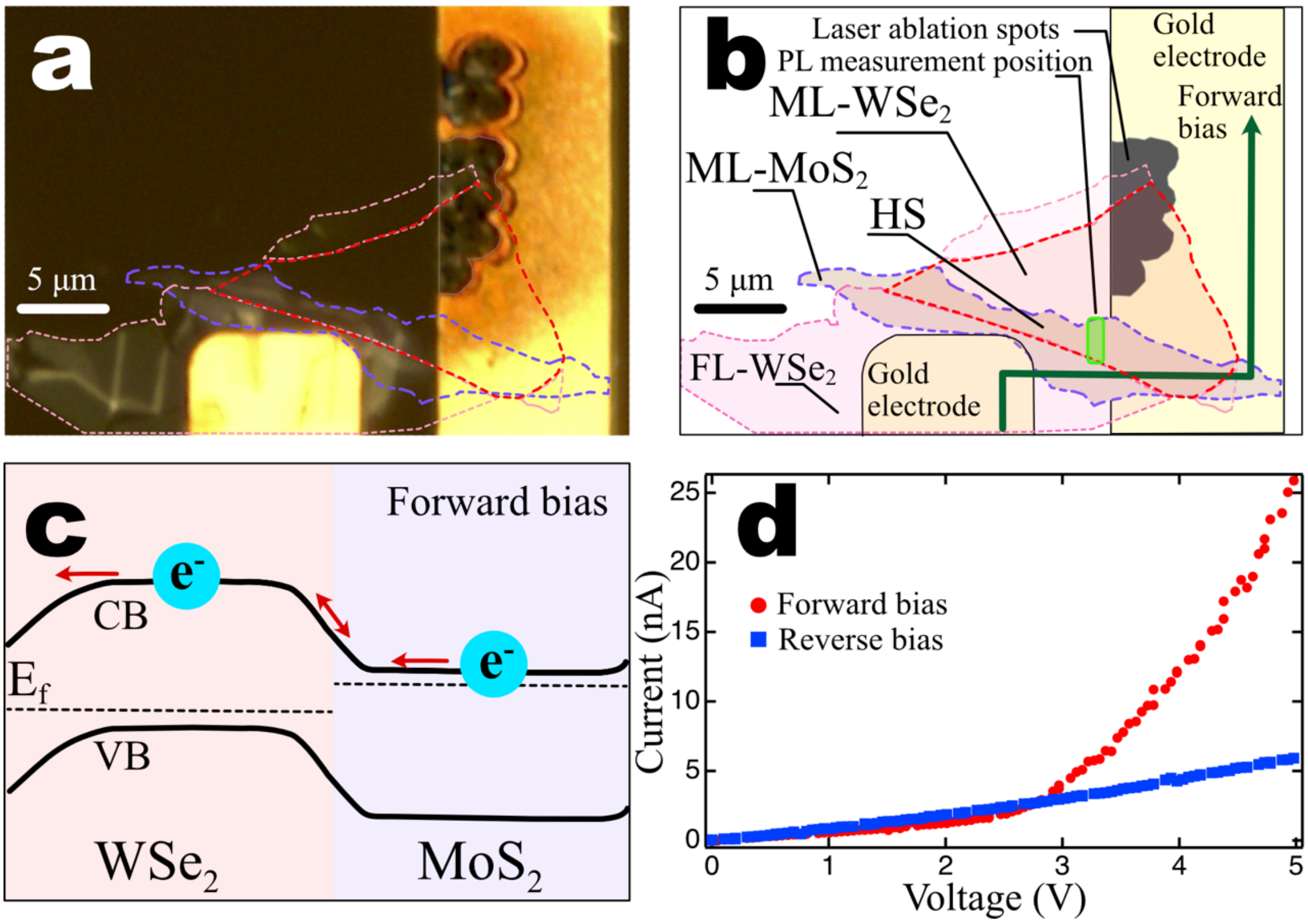

**Figure 1.**

(a) Optical micrograph of the fabricated $WSe_2/MoS_2$ heterostructure device. The dashed outlines indicate the monolayer (ML) and few-layer (FL) regions of the $MoS_2$ and $WSe_2$ flakes, as well as their overlap region (heterostructure, HS). Scale bar: 5 μm.

(b) Schematic illustration of the device configuration. The heterostructure was formed by stacking monolayer $WSe_2$ on monolayer $MoS_2$ on a commercial comb-shaped Au electrode. Unwanted bulk regions and excess $WSe_2$ were removed by laser ablation to isolate the transport path across the heterostructure. The gray regions indicate the positions of laser irradiation. The actual laser-affected region extends beyond the visibly damaged Au areas owing to the finite laser spot size. This is consistent with the discolored regions surrounding the laser-ablated spots visible in Fig. 1(a). The green rectangle indicates the position where the PL spectra were acquired. Optical measurements were performed prior to the removal process.

(c) Schematic band alignment of the $WSe_2/MoS_2$ heterostructure under forward bias. Under forward bias, electrons are injected from $MoS_2$ into $WSe_2$ across the heterointerface. Near the interface, electrons may undergo scattering before being extracted toward the $WSe_2$ electrode. CB: conduction band; VB: valence band.

(d) Current–voltage (*I*–*V*) characteristics of the device, exhibiting rectifying behavior consistent with diode operation across the $WSe_2/MoS_2$ heterostructure. Forward bias is defined as the condition where the $WSe_2$ side is positively biased relative to $MoS_2$, whereas reverse bias corresponds to the opposite polarity.

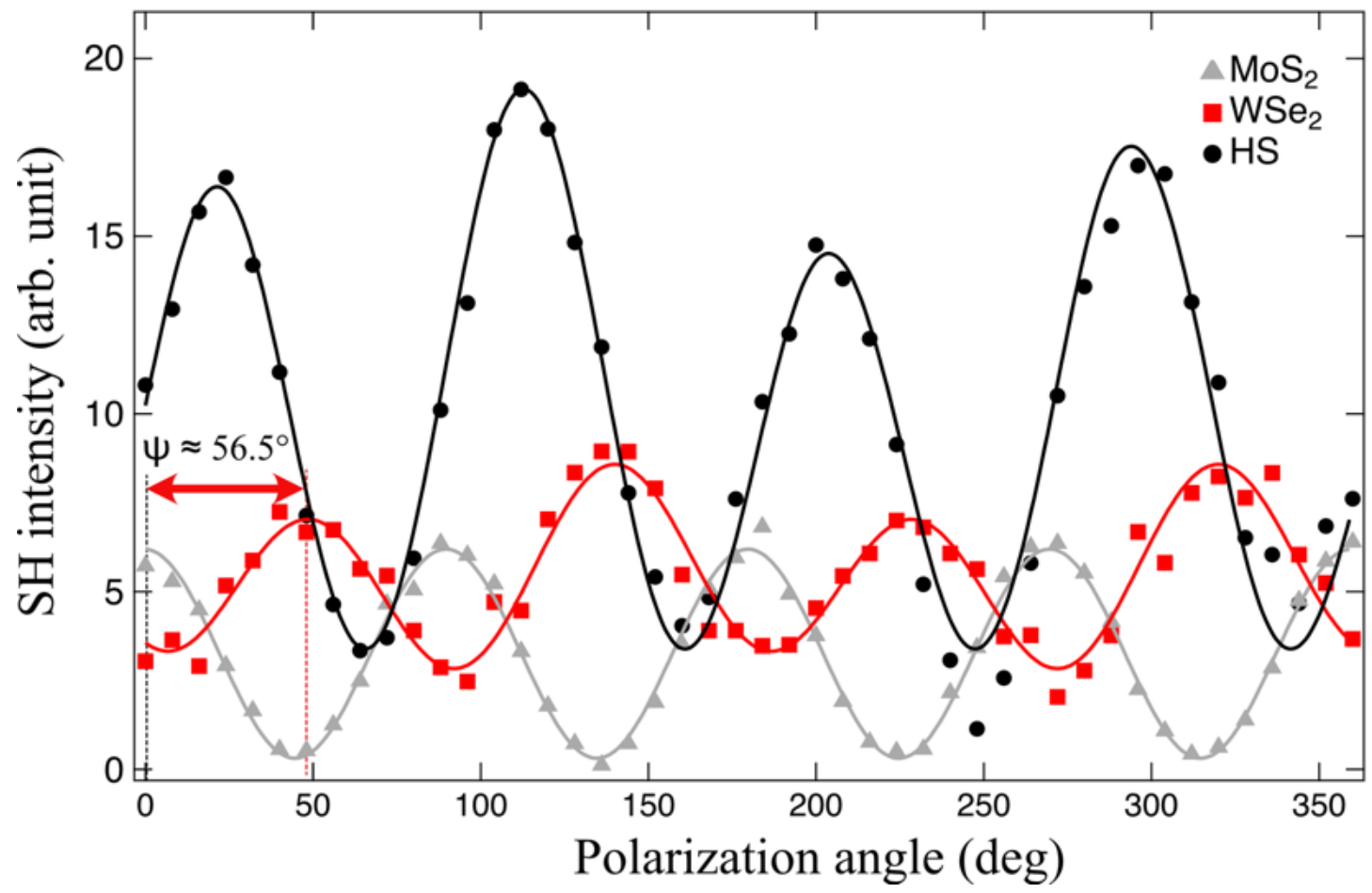

**Figure 2.** Incident polarization-resolved SHG signals from $MoS_2$, $WSe_2$, and the heterostructure (HS). The relative crystal orientation between $MoS_2$ and $WSe_2$ corresponds to a stacking angle of $\psi \approx 56.5°$, as determined from the fitted SHG phase. The SHG response from the HS appears between the two monolayer orientations, consistent with the formation of the stacked heterostructure.

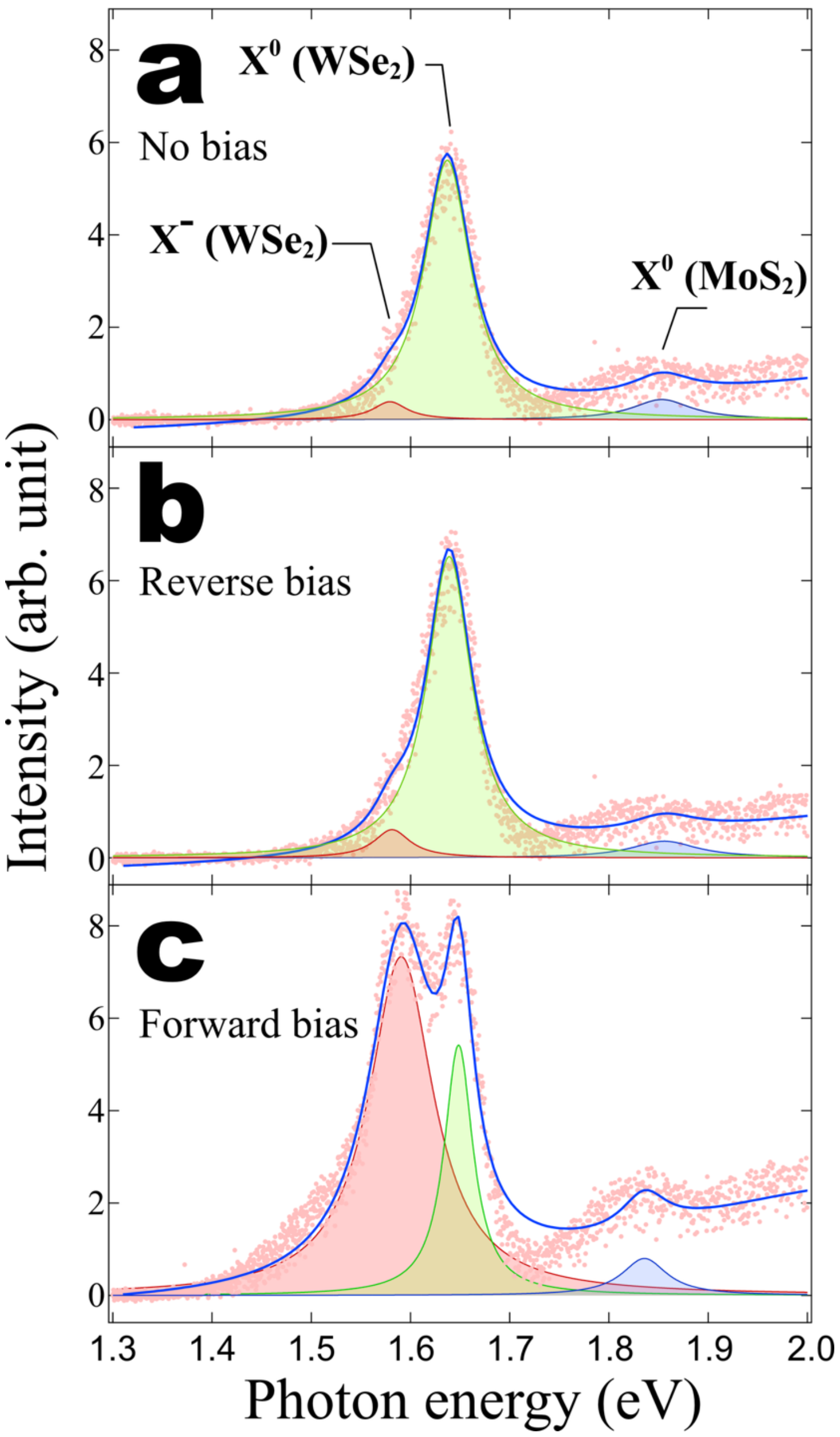

**Figure 3**. Bias-dependent photoluminescence spectra of the $WSe_2/MoS_2$ heterostructure.

(a–c) Photoluminescence spectra measured under (a) no bias, (b) reverse bias (−5 V), and (c) forward bias (+5 V). The spectra were fitted using multiple Lorentzian components corresponding to the neutral exciton $X^0$ and negatively charged trion $X^-$ of $WSe_2$, together with the neutral exciton emission from $MoS_2$. Under forward bias, the trion emission from $WSe_2$ is significantly enhanced. Multiple-Lorentzian spectral decomposition reveals an increase in the trion contribution accompanied by a reduction of the neutral exciton component, consistent with electron injection from $MoS_2$ into $WSe_2$ across the heterointerface.